\documentclass[preprint,showpacs,amsmath,nofootinbib]{revtex4}
\usepackage{color}

\definecolor{lightgray}{rgb}{.7,.7,.7}

\definecolor{red}{rgb}{1,0,0}

\definecolor{green}{rgb}{0,1,0}

\definecolor{blue}{rgb}{0,0,1}

\begin{document}

\title{Thermodynamics on the Maximally Symmetric Holographic Screen and Entropy from
Conical Singularities}

\author{Yu Tian}
\affiliation{College of Physical Sciences, Graduate University of
Chinese Academy of Sciences, Beijing 100049, China}
\affiliation{Kavli Institute for Theoretical Physics China, CAS,
Beijing 100190, China}

\author{Xiao-Ning Wu}
\affiliation{Institute of Mathematics, Academy of Mathematics and
System Science, CAS, Beijing 100190, China} \affiliation{Hua
Loo-Keng Key Laboratory of Mathematics, CAS, Beijing 100190, China}

\date{\today}

\begin{abstract}
For a general maximally symmetric (spherically, plane or hyperbola
symmetric) holographic screen, we rewrite the equations of motion of
general Lovelock gravity into the form of some generalized first law
of thermodynamics, under certain ansatz. With this observation
together with other two independent ways, exactly the same
temperature and entropy on the screen are obtained. So it is argued
that the thermodynamic interpretation of gravity is physically
meaningful not only on the horizon, but also on a general maximally
symmetric screen. Moreover, the formula of entropy is further
checked in the (maximally symmetric) general static case and
dynamical case. The entropy formula also holds for those cases.
Finally, the method of conical singularity is used to calculate the
entropy on such screen, and the result again confirms the entropy
formula.
\end{abstract}


\maketitle

\section{Introduction}

It is well known that black-hole thermodynamics \cite{thermo}
reveals a deep and elegant relation between gravity and
thermodynamics. This relation is also an important clue for
searching a quantum theory of gravity. In 1995, Jacobson found that
the Einstein equations can be obtained from the first law of
thermodynamics if assuming the area law of entropy for all local
acceleration horizons \cite{Jacobson}. This idea has been extended
to non-Einstein gravity theories (for a review, see \cite{review}).
A highlight on this problem recently is the idea of entropy force
proposed by Verlinde \cite{Verlinde}. Using the equipartition rule
of energy and the holographic principle
\cite{holographic,holography}, he found that gravity could be
understood as a kind of entropy force. This is a very attractive
idea both in physics and mathematics. On physical side, it gives a
new viewpoint of gravity in terms of holographic principle. On
mathematical side, this idea is closely related with initial
boundary value problem of Einstein equations. At almost the same
time, Padmanabhan reinterpreted the relation $E=2T S$
\cite{Padmanabhan1} between the Komar energy, temperature and
entropy as the equipartition rule of energy \cite{Padmanabhan2}.
Many research works have been done in this area (see
\cite{following,another,other,Tian,Chen,Cai,WGZY,CLW,Fursaev2} for
an incomplete list of them).

All these recent works imply that geometric quantities on a general
holographic screen also have thermodynamic interpretation as on a
black-hole horizon. As pointed out by Verlinde, for static case, an
Unruh-Verlinde temperature \cite{Verlinde} can be defined on a
general holographic screen. The relation between equipartition rule
of energy and gravitational field equation strongly supports that
such a temperature is a physical temperature. In \cite{Wald}, Wald
suggested a general method to get the horizon entropy and first law
of black hole for a general diffimorphism invariant theory.
Padmanabhan generalized this method to the off-horizon case in a
certain class of gravitational theories and also got one quarter of
the screen's area in Einstein's gravity \cite{another}. Many other
authors have also suggested $S=A/4$ for some general holographic
screen from different aspects \cite{other}, for example Fursaev's
result in the case of minimal surfaces \cite{Fursaev2}. (See,
however, \cite{Tian} for a different entropy formula, as will be
discussed later.) In Verlinde's derivation, he related the Komar
mass of the screen to the thermal energy of the screen, then got the
Einstein equations. A natural question is whether we can use other
types of quasi-local energy on the screen besides the Komar mass. In
ordinary thermodynamics, there are many types of energy, such as
inner energy, free energy, Gibbs free energy, $\cdots$, which
correspond to different thermal processes. So, it is reasonable to
guess that different quasi-local energy corresponds to different
kind of thermal energy, and all the hints of thermodynamics on an
off-horizon screen should fit into a whole picture. Chen et al have
made an attempt to this direction in four dimensional Einstein's
gravity and obtained a generalized first law of thermodynamics for
the spherically symmetric screen \cite{Chen}, but the energy
appearing there is the Arnowitt-Deser-Misner (ADM) mass instead of
some quasi-local energy associated to the screen. Even earlier, Cai
et al have also considered the spherically symmetric case in
Einstein's gravity and obtained a relation similar to the
generalized first law \cite{Cai} (see e.g. \cite{on-horizon} for the
on-horizon case), but that is a dynamical process, while in the
usual thermodynamic sense the generalized first law should describe
the quasi-static processes. Although most of the present works
support some thermodynamic relations on a general screen with
$S=A/4$ in Einstein's gravity, this entropy formula seems too simple
to be falsified. Therefore, it is necessary to investigate more
general theories of gravity, where expressions of the entropy and
other quantities are complicated enough, and to collect more,
different evidence for supporting that conclusion.

The aim of this paper is to consider more general gravity theory in
order to find more evidence to support the thermal interpretation of
gravity on an off-horizon screen. We consider a general spherical
screen in the Lovelock gravity in arbitrary dimensions
\cite{Lovelock} which is a natural generalization of Einstein's
gravity. Three independent tests support the thermal interpretation
in this case. First, we find that we can recast the equations of
motion into the form of the first thermodynamical law. From this
result, we can read out the quasi-local energy, entropy, temperature
associated to the screen directly. We find the quasi-local energy is
just the famous Misner-Sharp-like energy. Second, the analysis in
\cite{Chen} is generalized to this case, which for the
Reissner-Nordstr\"{o}m solution {in Einstein's gravity} involves the
Tolman-Komar energy inside the screen after a Legendre
transformation. Finally, we also find that the entropy obtained
previously just agrees with Padmanabhan's general definition of
entropy on the screen, which satisfies some equipartition-like rule.
In all these aspects, exactly the same entropy and Unruh-Verlinde
temperature arise, so it is convincing that the quantities and
thermodynamic interpretations on the screen are physically
meaningful.

Moreover, the formula of entropy is further confirmed in the general
static spherically symmetric case and dynamical spherically
symmetric case, as well as the corresponding plane symmetric and
hyperbola symmetric case in parallel. In those cases, all the
methods available also get exactly the same result of entropy, but
the forms of temperature are slightly different, on which we will
give some discussions.

Besides of those results, we also use another independent method to
check the entropy formula. It is well known that the entropy of
black hole horizon can be calculated by the method of conical
singularity \cite{resolve}. Such method have also been used to
calculate the holographic entanglement entropy \cite{HEE,complex}.
We apply this method to some general screen\footnote{We mean even
not necessarily a maximally symmetric screen. See the (Euclidean)
metric ansatz (\ref{general-dynamic}).} and find that the entropy
obtained by this method agrees with Padmanabhan's general definition
of off-horizon entropy (and also what we obtain from previous
methods for the maximally symmetric cases). This can be viewed as
independent evidence which supports the entropy formula.

This paper is organized as follows. In section II, we illustrate the
three methods in the Einstein case for static space-times with
spherical symmetry and the metric ansatz $g_{tt}=g^{-1}_{rr}$. In
section III, we generalize those results into the general Lovelock
gravity. The general static case is discussed in section IV. In this
section, we also discuss the non-stationary spherical solutions and
find that our results still hold in those cases. We also find that
our results hold if the spacetime is plane symmetric or hyperbola
symmetric. These cases is included in section V. The method of
conical singularity is discussed in section VI. In the last section,
we give some remarks and discussions on our results.

\section{Einstein's gravity}

Take {Einstein's gravity} in $n$ space-time dimensions as the
simplest example to illustrate our basic strategy. The action
functional is
\begin{equation}
I=\int d^n x (\frac{\sqrt{-g}}{16\pi} R+\mathcal{L}_\mathrm{matt}),
\end{equation}
which leads to the equations of motion
\begin{equation}\label{EOM}
R_{ab}-\frac 1 2 R g_{ab}=8\pi T_{ab}
\end{equation}
with $T^{ab}=\frac{2}{\sqrt{-g}}\frac{\delta}{\delta
g_{ab}}\int\mathcal{L}_\mathrm{matt} d^n x$ the stress-energy tensor
of matter. The most general static, spherically symmetric metric can
be written as
\begin{equation}\label{general-static}
ds^2=-h(r) dt^2+f(r)^{-1} dr^2+r^2 d\Omega_{n-2}^2
\end{equation}
with $d\Omega_{n-2}^2$ the metric on the unit $(n-2)$-sphere. We
will consider this general case, as well as the dynamical case, in
Section \ref{dynamical}. Here we assume the ansatz
\begin{equation}\label{ansatz}
ds^2=-f(r) dt^2+f(r)^{-1} dr^2+r^2 d\Omega_{n-2}^2
\end{equation}
for the metric, which means that the Lagrangian density
$\mathcal{L}_\mathrm{matt}$ of matter cannot be too arbitrary, while
still containing many cases of physical interest, such as
electromagnetic fields, the cosmological constant and homogeneous
ideal fluids \cite{fluids}, etc. In fact, the above ansatz
essentially requires the relation $T_t^t=T_r^r$ between the
components of the stress-energy tensor of matter. In this
space-time, the Unruh-Verlinde temperature on the spherical screen
of radius $r$ is easily obtained as
\begin{equation}\label{Unruh}
T=\frac{-\partial_r g_{tt}}{4\pi\sqrt{-g_{tt}
g_{rr}}}=\frac{f'}{4\pi},
\end{equation}
which is purely geometric and so independent of the gravitational
dynamics. Here a prime means differentiation with respect to $r$.

Upon substitution of the ansatz (\ref{ansatz}) into the equations of
motion (\ref{EOM}), the nontrivial part of them is
\cite{Padmanabhan}
\begin{equation}\label{non-trivial}
r f'-(n-3)(1-f)=\frac{16\pi P}{n-2} r^2
\end{equation}
with $P=T^r_r=T^t_t$ the radial pressure. Now we focus on a
spherical screen with fixed $f$ \cite{Chen} in different static,
spherically symmetric solutions of (\ref{EOM}). In order to do so,
we just need to compare two such configurations of infinitesimal
difference. In fact, multiplying both sides of (\ref{non-trivial})
by the factor
\begin{equation}
\frac{n-2}{16\pi}\Omega_{n-2} r^{n-4} dr,
\end{equation}
we have after some simple algebra (assuming $f$ fixed)
\begin{eqnarray}
&&\frac{f'}{4\pi} d\left(\frac{\Omega_{n-2}
r^{n-2}}{4}\right)-d\left(\frac{n-2}{16\pi}\Omega_{n-2} (1-f)
r^{n-3}\right)\nonumber\\
&=&P d(\frac{\Omega_{n-2} r^{n-1}}{n-1}).
\end{eqnarray}
The above equation is immediately recognized as the generalized
first law
\begin{equation}\label{first_law}
T dS-dE=PdV
\end{equation}
with $T$ the Unruh-Verlinde temperature (\ref{Unruh}) on the screen,
\begin{eqnarray}
S&=&\frac{\Omega_{n-2} r^{n-2}}{4},\label{entropy}\\
E&=&\frac{n-2}{16\pi}\Omega_{n-2} (1-f) r^{n-3}\label{energy}
\end{eqnarray}
and $V=\frac{\Omega_{n-2} r^{n-1}}{n-1}$ the volume of the
(standard) unit $(n-1)$-ball. Here $E$ is just the standard form of
the Misner-Sharp energy inside the screen in spherically symmetric
space-times \cite{M-S}, which is also identical to the
Hawking-Israel energy in this case. More explicitly, solving $f$
from (\ref{energy}) gives
\begin{equation}
f=1-\frac{16\pi E}{(n-2)\Omega_{n-2} r^{n-3}},
\end{equation}
which is the Schwarzschild solution in $n$ dimensions for constant
$E$ as its ADM mass, and some general spherically symmetric solution
for certain mass function $E(r)$.

Some remarks are in order. First, the generalized first law
(\ref{first_law}) is of the same form as that in \cite{Padmanabhan}
for the horizon of spherically symmetric black holes, but is valid
for general spherical screen with fixed $f$, which includes the
horizon as the special case $f=0$. Second, the entropy
(\ref{entropy}) in the generalized first law is actually $S=A/4$,
i.e. one quarter of the area, for a general spherical screen, the
same as the result obtained in \cite{Chen} by the generalized
Smarr's approach for the four dimensional case. (Similar results
appear in \cite{Cai} and \cite{another}, as mentioned previously.)

In fact, the generalized Smarr's approach can be used in the higher
dimensional case without any difficulty. The Reissner-Nordstr\"{o}m
solution in $n$ dimensions is
\begin{equation}
f=1-\frac{2\mu}{r^{n-3}}+\frac{q^2}{r^{2n-6}},
\end{equation}
where the mass parameter $\mu$ is related to the ADM mass $M$ by
$\mu=\frac{8\pi M}{(n-2)\Omega_{n-2}}$. For fixed $f$, in order to
obtain a generalized first law of the form \cite{Chen}
\begin{equation}\label{ADM}
dM=T dS+\phi dq
\end{equation}
with $T$ the Unruh-Verlinde temperature (\ref{Unruh}) on the screen,
one just needs to notice that
\begin{equation}
df(r,M,q)=\frac{\partial f}{\partial r} dr+\frac{\partial
f}{\partial M} dM+\frac{\partial f}{\partial q} dq=0
\end{equation}
gives
\begin{equation}\label{Smarr}
dM=-\frac{f'}{4\pi}(\frac{\partial f}{\partial M})^{-1}\frac{4\pi
dr}{dS} dS-(\frac{\partial f}{\partial M})^{-1}\frac{\partial
f}{\partial q} dq.
\end{equation}
Comparing (\ref{ADM}) and (\ref{Smarr}), we see that
\begin{equation}
S=-4\pi\int (\frac{\partial f}{\partial M})^{-1}
dr=\frac{\Omega_{n-2} r^{n-2}}{4},
\end{equation}
which is the same as (\ref{entropy}), and
\begin{equation}\label{phi}
\phi=(\frac{\partial f}{\partial M})^{-1}\frac{\partial f}{\partial
q}=\frac{(n-2)\Omega_{n-2} q}{8\pi r^{n-3}}
\end{equation}
proportional to the electrostatic potential on the screen.

Furthermore, by straightforwardly working out {the Tolman-Komar
energy $K=M-\phi q$ inside the screen}, which is just a Legendre
transformation of $M$, we can obtain another generalized first law
\begin{equation}\label{Komar}
dK=T dS-q d\phi,
\end{equation}
which seems even more relevant to the holographic picture, since now
all the quantities are closely related to the screen, and the
Tolman-Komar energy $K$ satisfies the equipartition rule
\cite{Padmanabhan2,Verlinde}. In this case the Misner-Sharp energy
$E=M-\frac{\phi q}{2}$ is different from either the ADM mass $M$ or
the Tolman-Komar energy $K$, which is a general fact except for the
vacuum case. Anyway, exactly the same temperature $T$ and entropy
$S$ appear in different kinds of generalized first
laws\footnote{There is no obvious relation between (\ref{Komar}) [or
(\ref{ADM})] with (\ref{first_law}), as can be seen more clearly in
the discussion around (\ref{relation}) for the general Lovelock
gravity.} and other places such as \cite{Cai} and \cite{another},
which is strong evidence that the Unruh-Verlinde temperature
(\ref{Unruh}) and the entropy (\ref{entropy}) should make sense in
physics. This argument will be further confirmed in more general
cases below.

\section{The Lovelock gravity}

Now we consider the general Lovelock gravity. The action functional
is
\begin{equation}\label{Lovelock_action}
I=\int d^n x (\frac{\sqrt{-g}}{16\pi}\sum_{k=0}^m\alpha_k
L_k+\mathcal{L}_\mathrm{matt})
\end{equation}
with $\alpha_k$ the coupling constants and
\begin{equation}
L_k=2^{-k}\delta^{a_1 b_1\cdots a_k b_k}_{c_1 d_1\cdots c_k d_k}
R_{a_1 b_1}^{c_1 d_1}\cdots R_{a_k b_k}^{c_k d_k},
\end{equation}
where $\delta^{ab\cdots cd}_{ef\cdots gh}$ is the generalized delta
symbol which is totally antisymmetric in both sets of indices. Note
that $\alpha_0$ is proportional to the cosmological constant and
$L_1=R$. This theory has the nice feature that it is free of ghosts,
and some special cases of it arise naturally as the low-energy
effective theories of string models \cite{no-ghost}.

By the ansatz (\ref{ansatz}) again and extending the approach in
\cite{Wheeler} for the vacuum case to include
$\mathcal{L}_\mathrm{matt}$, the nontrivial part of the equations of
motion is
\begin{equation}\label{Lovelock_EOM}
\sum_k\tilde{\alpha}_k (\frac{1-f}{r^2})^{k-1}[ k r
f'-(n-2k-1)(1-f)]=\frac{16\pi P}{n-2} r^2,
\end{equation}
where
\begin{equation}
\tilde{\alpha}_0=\frac{\alpha_0}{(n-1)(n-2)},\quad\tilde{\alpha}_1=\alpha_1,\quad\tilde{\alpha}_{k>1}=\alpha_k\prod_{j=3}^{2k}(n-j).
\end{equation}
Now follow the strategy illustrated in the Einstein case.
Multiplying both sides of the above equation by the factor
\begin{equation}\label{factor}
\frac{n-2}{16\pi}\Omega_{n-2} r^{n-4} dr,
\end{equation}
we have after some simple algebra (assuming $f$ fixed)
\begin{eqnarray}
&&\frac{f'}{4\pi} d\left(\frac{n-2}{4}\Omega_{n-2} r^{n-2} \sum_k
\frac{\tilde{\alpha}_k k}{n-2k} (\frac{1-f}{r^2})^{k-1}\right)\nonumber\\
&&-d\left(\frac{n-2}{16\pi}\Omega_{n-2}
r^{n-1}\sum_k\tilde{\alpha}_k
(\frac{1-f}{r^2})^k\right)\nonumber\\
&=&P d(\frac{\Omega_{n-2} r^{n-1}}{n-1}).\label{explicit_1st_law}
\end{eqnarray}
Recalling that the Unruh-Verlinde temperature (\ref{Unruh}) is
independent of the gravitational dynamics, we again recognize the
above equation as the generalized first law (\ref{first_law}) with
\begin{eqnarray}
S&=&\frac{n-2}{4}\Omega_{n-2} r^{n-2} \sum_k \frac{\tilde{\alpha}_k
k}{n-2k} (\frac{1-f}{r^2})^{k-1},\label{Lovelock_entropy}\\
E&=&\frac{n-2}{16\pi}\Omega_{n-2} r^{n-1}\sum_k\tilde{\alpha}_k
(\frac{1-f}{r^2})^k.\label{Lovelock_energy}
\end{eqnarray}
Here $E$ can be interpreted as some generalization of the
Misner-Sharp (or Hawking-Israel) energy to the Lovelock gravity (for
certain special case, see \cite{Lovelock_M-S} for the discussion of
the Misner-Sharp energy). In fact, when $E=M$ is a constant,
(\ref{Lovelock_energy}) is just the algebraic equation (of arbitrary
degree) that $f$ satisfies for the vacuum case \cite{Wheeler}, with
$M$ the ADM mass. And when
\begin{equation}\label{RN-like}
E(r)=M-\frac{\phi(r) q}{2}
\end{equation}
with $\phi(r)$ given by (\ref{phi}), (\ref{Lovelock_energy}) gives
the Reissner-Nordstr\"{o}m-like solution with charge $q$, while the
Born-Infeld-like case corresponds to more complicated mass function
$E(r)$ \cite{Lovelock-RN}.

The entropy (\ref{Lovelock_entropy}) should be discussed further. On
the horizon, we have $f=0$, so (\ref{Lovelock_entropy}) just becomes
the well-known entropy of the Lovelock black hole
\cite{Lovelock_BH}. On a general spherical screen with fixed $f$,
the generalized Smarr's approach can be applied without knowing the
explicit form of $f$ (which is impossible in the general Lovelock
gravity) and still gives the generalized first law (\ref{ADM}) for
the Reissner-Nordstr\"{o}m-like solution (\ref{RN-like}), with
exactly the same temperature (\ref{Unruh}) and entropy
(\ref{Lovelock_entropy}). In fact, from (\ref{Lovelock_energy}) and
(\ref{RN-like}) we have
\begin{equation}
M=\frac{n-2}{16\pi}\Omega_{n-2}\left(r^{n-1}\sum_k\tilde{\alpha}_k
(\frac{1-f}{r^2})^k+\frac{q^2}{r^{n-3}}\right).
\end{equation}
For this case, (\ref{Smarr}) can be rewritten as
\begin{equation}
dM=-\frac{f'}{4\pi}\frac{\partial M}{\partial f}\frac{4\pi dr}{dS}
dS-\frac{\partial M}{\partial q} dq.
\end{equation}
Noticing that the entropy (\ref{Lovelock_entropy}) and energy
(\ref{Lovelock_energy}) satisfy
\begin{equation}\label{relation}
\frac{\partial S}{\partial r}=-4\pi\frac{\partial E}{\partial
f}=-4\pi\frac{\partial M}{\partial f},
\end{equation}
which is independent of the previous interpretation of
(\ref{explicit_1st_law}) as the generalized first law
(\ref{first_law}), we see that (\ref{ADM}) really holds with $T$ in
(\ref{Unruh}), $S$ in (\ref{Lovelock_entropy}) and $\phi$ in
(\ref{phi}). {Actually, the first equality of (\ref{relation}) just
means that (\ref{factor}) is the integrating factor of the left hand
side of (\ref{Lovelock_EOM}), and this integrating factor just
renders the right hand side of (\ref{Lovelock_EOM}) to have the form
$P dV$, which is an interesting feature of the general Lovelock
gravity.

However, if we define the ``Tolman-Komar energy" $\tilde{K}=M-\phi
q$ and write down a generalized first law (\ref{Komar}) mimicking
the Einstein case, it is not clear whether $\tilde{K}$ has the
meaning of some quasi-local energy inside the screen. Instead, there
is a more acceptable Tolman-Komar energy \cite{another}, defined as
\begin{equation}
K=\int_\mathcal{V} d^{n-1}
x\frac{\sqrt{h}}{16\pi}\mathcal{R}^a_b\xi^b n_a
\end{equation}
for a region $\mathcal{V}$ with induced metric $h_{ab}$ and normal
vector $n^a$, where $\xi^b$ is the Killing vector and
\begin{equation}
\mathcal{R}^a_b=16\pi P^{acde} R_{bcde}
\end{equation}
with
\begin{equation}\label{P}
P^{abcd}=\frac{\partial L}{\partial R_{abcd}}.
\end{equation}
{Here $\mathcal{R}^a_b$ can be viewed as the generalization of Ricci
tensor to the Lovelock gravity.} Note that even for the vacuum case
the equations of motion do not imply $\mathcal{R}_{ab}=0$, so $K$
does not vanish, unlike in Einstein's gravity. In our case, we take
$\xi=\partial_t$ and $\mathcal{V}$ to be the outside of the screen.
If the space-time is asymptotically flat, it is natural to identify
\begin{equation}\label{Lovelock_Komar}
K=M-\int_\mathcal{V} d^{n-1}
x\frac{\sqrt{h}}{16\pi}\mathcal{R}^a_b\xi^b n_a
\end{equation}
as the Tolman-Komar energy inside the screen, where $M$ is the ADM
mass of the space-time. So, unlike in Einstein's gravity, this
energy is screen-dependent even in the vacuum case. For a
spherically symmetric screen in the RN-like solution
(\ref{RN-like}), the Tolman-Komar energy (\ref{Lovelock_Komar}) can
be explicitly computed to be
\begin{equation}\label{Tolman-Komar}
K=(n-2)^2\Omega_{n-2} \sum_k \frac{\tilde{\alpha}_k k}{n-2k}
(\frac{1-f}{r^2})^{k-1}\frac{r^{2n-4}\sum_{k}\tilde{\alpha}_{k}(n-2k-1)(\frac{1-f}{r^{2}})^{k}-(n-3)q^{2}}{16\pi
(n-3) r^{n-3}\sum_{k}\tilde{\alpha}_{k}k(\frac{1-f}{r^{2}})^{k-1}}.
\end{equation}
It is obvious that $K\ne M-\phi q$ in generic case, so it is not
clear whether a generalized first law concerning this Tolman-Komar
energy exists.}

Furthermore, Padmanabhan has proposed another definition of entropy
off the horizon \cite{another} in a certain class of theories
including the Lovelock gravity, generalizing the definition of
entropy on the horizon by Wald et al \cite{Wald}. For a general
screen $\mathcal{S}$, the associated entropy is suggested to be
\begin{equation}\label{general}
S=\int_\mathcal{S} {8}\pi
P^{ab}_{cd}\epsilon_{ab}\epsilon^{cd}\sqrt{\sigma} d^{n-2} x,
\end{equation}
$\epsilon_{ab}$ the binormal to $\mathcal{S}$ and $\sigma_{ab}$ the
metric on $\mathcal{S}$, where $L=\frac{1}{16\pi}\sum_k\alpha_k L_k$
for the Lovelock gravity. This entropy is shown to satisfy the
equipartition-like rule with the Unruh-Verlinde temperature
(\ref{Unruh}) and some generalized Tolman-Komar energy
\cite{another}. In our case, the only non-vanishing components of
the binormal are $\epsilon_{tr}=1/2=-\epsilon_{rt}$, so the only
relevant component of (\ref{P}) in (\ref{general}) is
\begin{equation}\label{Ptrtr}
P^{tr}_{tr}{=\frac{1}{16\pi}}\sum_k\alpha_k k 2^{-k}\delta^{tr a_2
b_2\cdots a_k b_k}_{tr c_2 d_2\cdots c_k d_k} R_{a_2 b_2}^{c_2
d_2}\cdots R_{a_k b_k}^{c_k d_k}
\end{equation}
with the indices $a_i,b_i,c_i,d_i$ ($i=2,\cdots,k$) running only
among the angular directions. By explicitly working out
\begin{equation}\label{R}
R^{ab}_{cd}=\frac{1-f}{r^2}\delta^{ab}_{cd}
\end{equation}
for the metric (\ref{ansatz}) and substituting it into
(\ref{Ptrtr}), one can see that {the general definition
(\ref{general}) of entropy} eventually gives
(\ref{Lovelock_entropy}). {In fact, substitution of (\ref{R}) into
(\ref{Ptrtr}) gives
\begin{eqnarray}
P^{tr}_{tr}&=&\frac{1}{16\pi}\sum_k\alpha_k k 2^{-k}\delta_{t r
c_{2} d_{2}\cdots c_{k} d_{k}}^{t r a_{2} b_{2}\cdots
a_{k} b_{k}}\delta_{a_{2} b_{2}}^{c_{2} d_{2}}\cdots\delta_{a_{k} b_{k}}^{c_{k} d_{k}}(\frac{1-f}{r^2})^{k-1}\nonumber\\
&=&\frac{1}{32\pi}\sum_k\alpha_k k\delta_{tra_{2}b_{2}\cdots
a_{k}b_{k}}^{tra_{2}b_{2}\cdots
a_{k}b_{k}}(\frac{1-f}{r^2})^{k-1}\nonumber\\
&=&\frac{1}{32\pi}\sum_k\alpha_k k
(n-2)(n-3)\cdots(n-2k+1)(\frac{1-f}{r^2})^{k-1}\nonumber\\
&=&\frac{n-2}{32\pi}\sum_k\frac{\tilde{\alpha}_k k}{n-2k}
(\frac{1-f}{r^2})^{k-1}
\end{eqnarray}
Substituting the above expression into (\ref{general}), we then
obtain the entropy (\ref{Lovelock_entropy}) exactly. Moreover, the
entropy (\ref{Lovelock_entropy}), the Unruh-Verlinde temperature
(\ref{Unruh}) and the Tolman-Komar energy (\ref{Tolman-Komar})
satisfy
\begin{equation}
K=\frac{n-2}{n-3} T S,
\end{equation}
as Padmanabhan proved.}

\section{The general static case and dynamical case}
\label{dynamical}

For the general static case in the Lovelock gravity, it is
convenient to let $h(r)=e^{-2c(r)} f(r)$ in the metric
(\ref{general-static}). In this case, the $tt$ and $rr$ components
\begin{eqnarray}
  \sum_k\tilde{\alpha}_k (\frac{1-f}{r^2})^{k-1}[ k r
f'-(n-2k-1)(1-f)] &=& \frac{16\pi T_t^t}{n-2} r^2,\label{Gtt}\\
  \sum_k\tilde{\alpha}_k (\frac{1-f}{r^2})^{k-1}[ k r
(f'-2f c')-(n-2k-1)(1-f)] &=& \frac{16\pi T_r^r}{n-2} r^2\label{Grr}
\end{eqnarray}
of the equations of motion are independent of each other. {Focusing
on a screen with fixed $f$,} taking a linear combination of these
two equations and multiplying both sides of the equation by the
factor (\ref{factor}), we have again {the generalized first law}
(\ref{first_law}) with exactly the same entropy
(\ref{Lovelock_entropy}) and Misner-Sharp-like energy
(\ref{Lovelock_energy}), but with a slightly different temperature
\begin{equation}\label{gen_temperature}
T=\frac{f'-2l f c'}{4\pi}=\frac{\partial_r [(g^{rr})^{1-l}
(-g_{tt})^l]}{4\pi (-g_{tt} g_{rr})^l}
\end{equation}
from (\ref{Unruh}) and
\begin{equation}\label{gen_pressure}
P=(1-l) T_t^t+l T_r^r.
\end{equation}
Now, two special cases should be noted. The first one is the choice
$l=1/2$, where $P=(T_t^t+T_r^r)/2$ and it can be shown that
\begin{equation}
T=\frac{e^{c} (e^{-c}f)'}{4\pi}=\frac{D_a D^a r}{4\pi}\qquad (a=t,r)
\end{equation}
just coincides with Hayward's definition \cite{Hayward} generalized
to the off-horizon case \cite{Cai} in Einstein's
gravity.\footnote{The work density $w$ there is just $-P$ here.} The
second one is the choice $l=1$, which is of particular interest,
since in this case $P=T_r^r$ is just the standard expression of the
radial pressure and
\begin{equation}\label{mod_temperature}
T=\frac{\partial_r g_{tt}}{4\pi g_{tt} g_{rr}}
\end{equation}
differs from the standard Unruh-Verlinde temperature only by a
$\sqrt{-g_{tt} g_{rr}}$ factor. {Similar phenomena of non-unique
temperatures are extensively observed in the on-horizon case
\cite{non-unique,Cai}.

For the general definition (\ref{general}) of entropy, since
(\ref{R}) still holds in this case and it is easily seen that
\begin{equation}\label{epsilon^2}
\epsilon_{tr}\epsilon^{tr}=\frac 1 4
\end{equation}
for the metric (\ref{general-static}), the calculation in the
previous section leads to exactly the same result
(\ref{Lovelock_entropy}), which is consistent with the above
analysis that leads to the generalized first law.}

The dynamical case is formally the same. In this case, the metric
can be written as
\begin{equation}\label{dynamical_metric}
ds^2=-e^{-2c(t,r)} f(t,r) dt^2+f(t,r)^{-1} dr^2+r^2 d\Omega_{n-2}^2,
\end{equation}
with $t$ the so-called Kodama time \cite{Kodama}. It turns out that
the above discussion extends to this case straightforwardly, with
$f$, $c$ and other quantities in the same equations
(\ref{Gtt})-(\ref{mod_temperature}) understood as functions of both
$t$ and $r$. However, the thermodynamic meaning of the resulting
``first law"
\begin{equation}\label{fake_1st_law}
T dS-dE=PdV
\end{equation}
is obscure. In fact, we know in the standard thermodynamics that the
expression $dQ=T dS$ holds only in reversible processes, and a
reversible process is necessarily quasi-static. Thus, for such a
dynamical evolving case, (\ref{fake_1st_law}) cannot be regarded as
the generalized first law in the usual thermodynamic sense.
Nevertheless, the first-law-like description (\ref{fake_1st_law}) of
the Lovelock gravitational dynamics is interesting in its own right.
And needless to say, exactly the same, unique expression
(\ref{Lovelock_entropy}) of entropy on the screen involved here
further confirms its physical significance. {Furthermore, the
general definition (\ref{general}) of entropy still gives
(\ref{Lovelock_entropy}) even in this case, since (\ref{R}) and
(\ref{epsilon^2}) still hold for the metric
(\ref{dynamical_metric}).}

\section{The plane symmetric and hyperbola symmetric cases}

{The above discussions can be generalized to the plane symmetric and
hyperbola symmetric cases. Generally, the static metric can be
written as
\begin{equation}\label{symmetric_metric}
ds^2=-e^{-2c(r)} f(r) dt^2+f(r)^{-1} dr^2+r^2
d\Omega_{\varepsilon,n-2}^2
\end{equation}
with $d\Omega_{\varepsilon,n-2}^2$ the metric on the ``unit"
$(n-2)$-sphere, plane or hyperbola for $\varepsilon$ equal to $1$,
$0$ or $-1$, respectively. In these cases, (\ref{Gtt}) and
(\ref{Grr}) become
\begin{eqnarray}
  \sum_k\tilde{\alpha}_k (\frac{\varepsilon-f}{r^2})^{k-1}[ k r
f'-(n-2k-1)(\varepsilon-f)] &=& \frac{16\pi T_t^t}{n-2} r^2,\\
  \sum_k\tilde{\alpha}_k (\frac{\varepsilon-f}{r^2})^{k-1}[ k r
(f'-2f c')-(n-2k-1)(\varepsilon-f)] &=& \frac{16\pi T_r^r}{n-2} r^2.
\end{eqnarray}
Similar discussions as in the previous section lead to the
generalized first law (\ref{first_law}) with the entropy and
Misner-Sharp energy
\begin{eqnarray}
S&=&\frac{n-2}{4}\Omega_{n-2} r^{n-2} \sum_k \frac{\tilde{\alpha}_k
k}{n-2k} (\frac{\varepsilon-f}{r^2})^{k-1},\label{symmetric_entropy}\\
E&=&\frac{n-2}{16\pi}\Omega_{n-2} r^{n-1}\sum_k\tilde{\alpha}_k
(\frac{\varepsilon-f}{r^2})^k,\label{symmetric_energy}
\end{eqnarray}
while the temperature (\ref{gen_temperature}) and pressure
(\ref{gen_pressure}) have the same forms in all these cases.

On the horizon ($f=0$), these results just become the known ones for
the plane symmetric and hyperbola symmetric black holes in the
general Lovelock gravity. Especially, in the plane symmetric case
($\varepsilon=0$), the entropy on the horizon is just $A/4$ and the
corresponding Misner-Sharp energy is simply
\begin{equation}
\frac{n-2}{16\pi}\tilde{\alpha}_0\Omega_{n-2} r^{n-1}
\end{equation}
even in the general Lovelock gravity \cite{planar}, while it is
obvious that this statement no longer holds off the horizon.

For the dynamical case, it is easy to check that the above
discussion applies straightforwardly, with $f$, $c$ and other
quantities in the same equations
(\ref{symmetric_metric})-(\ref{symmetric_energy}) understood as
functions of both $t$ and $r$. Furthermore, the definition
(\ref{general}) of entropy again gives (\ref{symmetric_entropy}) in
the general static case and dynamical case, since (\ref{epsilon^2})
still holds and (\ref{R}) now becomes
\begin{equation}
R^{ab}_{cd}=\frac{\varepsilon-f}{r^2}\delta^{ab}_{cd}
\end{equation}
for the metric (\ref{symmetric_metric}), even with $f$ and $c$
understood as functions of both $t$ and $r$.}

\section{The entropy from conical singularities}

{In the Euclidean approach to the thermodynamics of black holes, the
horizon temperature for the metric (\ref{ansatz}) is determined by
requiring that its Euclidean counterpart
\begin{equation}
ds^2=f(r) d\tau^2+f(r)^{-1} dr^2+r^2 d\Omega_{n-2}^2
\end{equation}
should have no conical singularity at the origin $f(r)=0$ of the
$\tau$-$\rho$ plane, where $\rho$ is defined by the differential
equation
\begin{equation}
d\rho=f(r)^{-1/2} dr
\end{equation}
with the boundary condition $\rho=0$ when $f(r)=0$. Note that this
origin just corresponds to the horizon of the black hole. That
analysis results in a periodicity
\begin{equation}
\tau\sim\tau+\frac{4\pi}{f'(r_0)}\qquad (f(r_0)=0)
\end{equation}
of the imaginary time $\tau$ at the origin, which implies a
temperature
\begin{equation}
T=\frac{f'(r_0)}{4\pi}
\end{equation}
of the horizon.

On the other hand, if we allow a conical singularity at the origin
with an angle $\beta$, then the Euclidean action $I_\mathrm{E}$
(upon properly renormalized) will be $\beta$-dependent and,
correspondingly, there is an entropy
\begin{equation}\label{off-shell}
S=\lim_{\beta\to 2\pi} (\beta\frac{\partial}{\partial\beta}-1)
I_\mathrm{E}
\end{equation}
canonically conjugate to the deficit angle $\delta=2\pi-\beta$,
which is shown to agree with the black-hole entropy in the general
Lovelock gravity \cite{canonical}. Note that this viewpoint on the
entropy is off shell, i.e. does not rely on the validity of the
equations of motion, in contrast to the on-shell ones by the usual
black-hole thermodynamics or by the method of Wald et al
\cite{Wald}. An analysis of Riemann manifolds with conical
singularities, using some regularization techniques, has been made
in \cite{resolve}, under certain requirements on the asymptotic
behavior of the metric approaching the singularity. However, we will
use another approach \cite{complex} in the following to analyze the
conical singularity and then compute the entropy (\ref{off-shell}),
which seems more convenient and relevant to the off-horizon case
that we are really interested in throughout this paper.

In fact, we take a more general form
\begin{equation}\label{general-dynamic}
ds^{2}=g_{\tau\tau}(\tau,r) d\tau^2+2 g_{\tau r}(\tau,r) d\tau
dr+g_{rr}(\tau,r) dr^2+g_{ij}(\tau,r,x)dx^{i}dx^{j}
\end{equation}
of metric, which is assumed regular everywhere and can be
{(globally)} recast as
\begin{equation}\label{complex}
ds^{2}=2g_{w\bar{w}}(w,\bar{w})dw
d\bar{w}+g_{ij}(w,\bar{w},x)dx^{i}dx^{j}
\end{equation}
under some conformally flat complex coordinates $(w,\bar w)$ on the
$(\tau,r)$ plane. Here $i,j$ run over indices excluding $\tau$ and
$r$. The Euclidean spherically symmetric, plane symmetric and
hyperbola symmetric space-times are all special cases of the metric
(\ref{general-dynamic}). For the on-horizon, static case discussed
previously the origin $w=0=\bar w$ of the complex plane is just at
$f(r)=0$ (or $\rho=0$), while for the off-horizon, static case the
origin of the complex plane is at $(\tau,r)$ with $r$ the position
of the screen and $\tau$ arbitrary. The conically singular, or
multi-sheeted, structure is realized by the coordinate
transformation
\begin{equation}\label{multi-sheet}
w=z^m,
\end{equation}
which is singular at $z=0=\bar z$ for $m\ne 1$. {Precisely, our
original Euclidean space-time has topology $R^2\times\mathcal{S}$.
To obtain the multi-sheeted Euclidean space-time, $m$ copy of the
$R^2$ factor is glued together in the standard way as Riemann
surfaces \cite{complex}, while the $\mathcal{S}$ factor remains
untouched.} The metric (\ref{complex}) then becomes
\begin{equation}
ds^{2}=2m^2 (z\bar z)^{m-1} g_{w\bar{w}}dz
d\bar{z}+g_{ij}dx^{i}dx^{j}.
\end{equation}
Note that the conical angle $\beta=2\pi m$. Since $\beta\to 2\pi$ in
(\ref{off-shell}), we extend $m$ from integers to reals and let
$m=1+\epsilon$ with $\epsilon$ an infinitesimal parameter. To the
linear order of $\epsilon$, the above metric is
\begin{equation}
ds^{2}=2[1+2\epsilon+\epsilon\ln(z\bar z)] g_{w\bar{w}}dz
d\bar{z}+g_{ij}dx^{i}dx^{j}.
\end{equation}
Straightforward calculations give the Riemann curvature
\begin{equation}\label{singular}
{}^{\epsilon}R^{ab}_{cd}=R^{ab}_{cd}-8\pi\epsilon g^{z\bar
z}\epsilon^{ab}\epsilon_{cd}\delta^2(z,\bar z)
\end{equation}
with $R^{ab}_{cd}$ the tensorial part of the Riemann curvature,
$\epsilon^{ab}$ the binormal to the surface $z=0=\bar z$ and the
second term on the right hand side the non-tensorial part due to the
conical singularity, which is similar to the result in
\cite{resolve}.

Now suppose that the Euclidean action has the form
\begin{equation}\label{Euclidean}
-I_\mathrm{E}=\int F(g_{ef},R^{ab}_{cd},\psi)\sqrt{g} dw d\bar{w}
d^{n-2}x,
\end{equation}
where $F(g_{ef},R^{ab}_{cd},\psi)$ is some scalar function of the
metric, the Riemann curvature (and its contractions) and some matter
fields (and their covariant derivatives) collectively denoted by
$\psi$. Under the transformation (\ref{multi-sheet}), substitution
of (\ref{singular}) into (\ref{Euclidean}) gives the
$\beta$-dependent Euclidean action
\begin{eqnarray}
{}^{\epsilon}I_\mathrm{E}&=&\epsilon\int 8\pi\frac{\partial
F}{\partial R^{ab}_{cd}}\epsilon^{ab}\epsilon_{cd}\delta^2(z,\bar z)
g^{z\bar z}\sqrt{g} dz d\bar{z} d^{n-2}x-\int F\sqrt{g} dz
d\bar{z} d^{n-2}x\nonumber\\
&=&\epsilon\int 8\pi\frac{\partial F}{\partial
R^{ab}_{cd}}\epsilon^{ab}\epsilon_{cd}\sqrt{\sigma} d^{n-2}x-\int
F\sqrt{g} dz d\bar{z} d^{n-2}x,\label{multi-Euclidean}
\end{eqnarray}
where $\sigma$ is the determinant of the induced metric
$\sigma_{ij}$ on the surface $z=0=\bar z$ of codimension 2. Note
that the precise meaning of the bulk integral $\int\sqrt{g} dz
d\bar{z} d^{n-2}x$ in (\ref{multi-Euclidean}) is to evaluate it for
integral $m$ and extend the result to real $m\to 1$ \cite{Fursaev},
so this integral is by construction proportional to $\beta$.
Substituting (\ref{multi-Euclidean}) into (\ref{off-shell}), we then
have
\begin{equation}
S=\int 8\pi\frac{\partial F}{\partial
R^{ab}_{cd}}\epsilon^{ab}\epsilon_{cd}\sqrt{\sigma} d^{n-2}x,
\end{equation}
which agrees with (\ref{general}) because we know from
(\ref{Euclidean}) that $F$ is just the Euclidean version of the
Lagrangian $L$.} {Similar argument has early been made in
\cite{Nelson} that the horizon entropy from conical singularities is
equivalent to the definition by Wald et al, but introducing and
treating the singularities by techniques similar to \cite{resolve}.}

\section{Concluding remarks}

In this work, we use three independent ways to check the
thermodynamical interpretation of some geometric quantities on a
maximally symmetric (spherically, plane or hyperbola symmetric)
holographic screen in the general Lovelock theory, first under
certain metric ansatz in the static case. All these methods give the
same Unruh-Verlinde temperature, Misner-Sharp energy and entropy
formula on the screen. This agreement supports the thermal
interpretation of these geometric quantities to be physically
meaningful. In the general static spherical case and dynamical
spherical case, then, exactly the same form of entropy appears, but
the definition of temperature is of some ambiguity and the physical
meaning of the ``generalized first law" in the dynamical case is
obscure, which have been clarified in our paper. In fact, we have
obtained a series of ``generalized first law", which include that of
Hayward as a special case. Similar to the on-horizon case that the
entropy is canonically conjugate to the deficit angle of the conical
singularity at the ``horizon" of the Euclidean space-time, the
off-horizon entropy has been shown to be canonically conjugate to
the deficit angle of the conical singularity at the ``screen" from
the Euclidean point of view. This result can been viewed as
additional independent evidence of the thermal meaning of geometric
quantities on a general screen.

Nevertheless, there are many open questions and/or unclear points in
this framework, of which an important one will be described as
follows, simply in Einstein's gravity. Although the relation $S=A/4$
for a general spherically symmetric screen seems rather universal
and is supported by many recent works, there is an alternative
expression $S=2\pi R E$ obtained in \cite{Tian}, which seems also
substantial. {To get the former form of entropy, we have to focus on
a screen with fixed $f$, while to get the latter form, we have to
focus on a screen with fixed $r$.} In fact, the former form of
entropy just saturates the holographic entropy bound
\cite{holographic}, while the latter form just saturates the
Bekenstein entropy bound \cite{Bekenstein}. Furthermore, for the
former form of entropy it is easy to write down some generalized
first law of thermodynamics as discussed above, but it is not clear
how to realize Verlinde's entropy variation formula and then the
gravity as an entropic force, while for the latter form there exist
the entropy variation formula and the entropic force expression
\cite{Tian} but without a satisfactory generalized first law. How to
reconcile these two forms of entropy is a significant open question.

Although in the method of conical singularity, Padmanabhan's general
definition of off-horizon entropy has been confirmed in a much
larger class of metrics than the maximally symmetric ones, it seems
that it is difficult to generalize all the other approaches of
investigating the off-horizon entropy, and moreover, the off-horizon
thermodynamics to a general (non-maximally-symmetric) holographic
screen. This problem should be left for future works.

\begin{acknowledgments}
We thank Prof. R.-G. Cai, C.-G. Huang, Y. Ling, C.-J. Gao, Dr.
{J.-R. Sun} and H.-B. Zhang for helpful discussions. This work is
partly supported by the National Natural Science Foundation of China
(Grant Nos. 10705048, 10731080 and {11075206}) and the President
Fund of GUCAS.
\end{acknowledgments}


\begin{thebibliography}{99}

\bibitem{thermo} J.M. Bardeen, B. Carter and S.W. Hawking,
Commun. Math. Phys. 31, 161 (1973); J.D. Bekenstein, Phys. Rev. D 7,
949 (1973); J.D. Bekenstein, Phys. Rev. D 7, 2333 (1973); S.W.
Hawking, Commun. Math. Phys. 43, 199 (1975) [Erratum-ibid. 46, 206
(1976)].

\bibitem{Jacobson} T. Jacobson, Phys. Rev. Lett. 75, 1260 (1995).

\bibitem{review} T. Padmanabhan, Rep. Prog. Phys. 73 (2010) 046901 [arXiv:0911.5004].

\bibitem{Verlinde} E.P. Verlinde, [arXiv:1001.0785].

\bibitem{holographic} L. Susskind, J. Math. Phys. 36, 6377 (1995); G. 't Hooft, [gr-qc/9310026].

\bibitem{holography} E. Witten, Adv. Theor. Math. Phys. 2, 253 (1998).

\bibitem{Padmanabhan1} T. Padmanabhan, Class. Quantum Grav. 21, 4485 (2004) [gr-qc/0308070].

\bibitem{Padmanabhan2} T. Padmanabhan, Mod. Phys. Lett. A 25 1129 (2010) [arXiv:0912.3165].

\bibitem{following} F.-W. Shu and Y.-G. Gong, [arXiv:1001.3237];
R.-G. Cai, L.-M. Cao and N. Ohta, [arXiv:1001.3470]; Y. Zhang, Y.-G.
Gong and Z.-H. Zhu, [arXiv:1001.4677]; S.-W. Wei, Y.-X. Liu and
Y.-Q. Wang, [arXiv:1001.5238]; Y. Ling and J.-P. Wu,
[arXiv:1001.5324]; L. Smolin, [arXiv:1001.3668]; T. Wang,
[arXiv:1001.4965]; M. Li and Y. Wang, [arXiv:1001.4466]; J. Makela,
[arXiv:1001.3808]; F. Caravelli and L. Modesto, [arXiv:1001.4364];
J.-W. Lee, H.-C. Kim and J. Lee, [arXiv:1001.5445]; C.-J. Gao,
[arXiv:1001.4585]; J. Munkhammar, [arXiv:1003.1262]; L. Zhao,
[arXiv:1002.0488]; Y. Zhao, [arXiv:1002.4039]; X. Kuang, Y. Ling and
H. Zhang, [arXiv:1003.0195]; X.-G. He and B.-Q. Ma,
[arXiv:1003.1625]; X. Li and Z. Chang, [arXiv:1005.1169]; H. Wei,
[arXiv:1005.1445]; J.-W. Lee, [arXiv:1003.4464]; Ee Chang-Young, M.
Eune and K. Kimm, [arXiv:1003.2049]; Y.S. Myung, [arXiv:1002.0871];
Y.S. Myung and Y.-W. Kim, [arXiv:1002.2292]; Y.S. Myung,
[arXiv:1003.5037]; I.V. Vancea and M.A. Santos, [arXiv:1002.2454];
J. Kowalski-Glikman, [arXiv:1002.1035]; R.A. Konoplya,
[arXiv:1002.2818]; A. Sheykhi, [arXiv:1004.0627]; S. Samanta,
[arXiv:1003.5965]; C.M. Ho, D. Minic and Y.J. Ng, [arXiv:1005.3537];
M. Li and Y. Pang, [arXiv:1004.0877]; Y.-X. Liu, Y.-Q. Wang and
S.-W. Wei, [arXiv:1002.1062]; S. Ghosh, [arXiv:1003.0285]; P.
Nicolini, [arXiv:1005.2996]; R. Banerjee and B.R. Majhi, Phys. Rev.
D 81, 124006 (2010); W. Gu, M. Li and R.-X. Miao, [arXiv:1011.3419].

\bibitem{another} T. Padmanabhan, Phys. Rev. D 81, 124040 (2010) [arXiv:1003.5665].

\bibitem{other} Q. Pan and B. Wang, [arXiv:1004.2954]; V.V. Kiselev and S.A. Timofeev, [arXiv:1004.3418]; J.-P. Lee, [arXiv:1005.1347].

\bibitem{Tian} Y. Tian and X.-N. Wu, Phys. Rev. D 81, 104013 (2010) [arXiv:1002.1275].

\bibitem{Chen} Y.-X. Chen and J.-L. Li, [arXiv:1006.1442].

\bibitem{Cai} R.-G. Cai, L.-M. Cao and N. Ohta, Phys. Rev. D 81, 084012 (2010) [arXiv:1002.1136].

{\bibitem{WGZY} S.-F. Wu, X.-H. Ge, P.-M. Zhang and G.-H. Yang,
[arXiv:1008.3215].

\bibitem{CLW} Y.-X. Chen, J.-L. Li and Y.-Q. Wang,
[arXiv:1008.3215].}

\bibitem{Wald} R.M. Wald, Phys. Rev. D 48 (1993) 3427 [gr-qc/9307038]; V. Iyer and R.M. Wald, Phys. Rev. D 52 (1995) 4430 [gr-qc/9503052].

\bibitem{on-horizon} R.-G. Cai and L.-M. Cao, Phys. Rev. D 75, 064008 (2007) [gr-qc/0611071].

\bibitem{Lovelock} D. Lovelock, J. Math. Phys. 12, 498 (1971).

\bibitem{non-Einstein} R.-G. Cai and N. Ohta, Phys. Rev. D 81, 084061 (2010);
R.-G. Cai, L.-M. Cao, Y.-P. Hu and S.P. Kim, Phys. Rev. D 78, 124012
(2008); R.-G. Cai, Phys. Rev. D 65, 084014 (2002); R.-G. Cai and
K.-S. Soh, Phys. Rev. D 59, 044013 (1999); R.-G. Cai and Qi Guo,
Phys. Rev. D 69, 104025 (2004).

{\bibitem{planar} R.-G. Cai, Phys. Lett. B 582, 237 (2004).}

\bibitem{fluids} V.V. Kiselev, Class. Quant. Grav. 20 (2003) 1187.

\bibitem{Padmanabhan} A. Paranjape, S. Sarkar and T. Padmanabhan, Phys. Rev. D 74, 104015 (2006) [hep-th/0607240].

\bibitem{M-S} C.W. Misner and D.H. Sharp, Phys. Rev. 136 (1964) B571.

\bibitem{no-ghost} B. Zwiebach, Phys. Lett. B 156 (1985) 315; B. Zumino, Berkeley preprint UCB-PTH-85/13, LBL-19302.

\bibitem{Wheeler} J.T. Wheeler, Nucl. Phys. B 273 (1986) 732.

\bibitem{Lovelock_M-S} H. Maeda and M. Nozawa, Phys. Rev. D 77, 064031
(2008) [arXiv:0709.1199].

\bibitem{Lovelock-RN} D. Wiltshire, Phys. Lett. B 169 (1986) 36; D. Wiltshire, Phys.
Rev. D 38 (1988) 2445.

\bibitem{Lovelock_BH} T. Jacobson and R. Myers, Phys. Rev. Lett. 70 (1993) 3684
[hep-th/9305016].

\bibitem{Hayward} S.A. Hayward, Class. Quant. Grav. 15, 3147 (1998) [gr-qc/9710089]; S.A. Hayward, S. Mukohyama and M.C. Ashworth, Phys. Lett. A 256, 347 (1999)
[gr-qc/9810006].

\bibitem{non-unique} S.A. Hayward et al, [arXiv:0806.0014]; R.-G. Cai, \textit{Gravity from Thermodynamics}, talk given at Peking
University (2010).

\bibitem{Kodama} G. Abreu and M. Visser, [arXiv:1004.1456].

\bibitem{canonical} M. Ba\~{n}ados, C. Teitelboim and J. Zanelli, Phys. Rev. Lett. 72, 957
(1994) [gr-qc/9309026].

\bibitem{resolve} D.V. Fursaev and S.N. Solodukhin, Phys. Rev. D 52 (1995)
2133 [hep-th/9501127].

\bibitem{HEE} S. Ryu and T. Takayanagi, Phys. Rev. Lett. 96 (2006)
181602 [hep-th/0603001]; S. Ryu and T. Takayanagi, JHEP 0608 (2006)
045 [hep-th/0605073]{; R.C. Myers and A. Sinha, [arXiv:1011.5819]}.

\bibitem{complex} A. Schwimmer and S. Theisen, Nucl. Phys. B 801, 1 (2008) [arXiv:0802.1017];
G. Michalogiorgakis, JHEP 0812, 068 (2008) [arXiv:0806.2661]; J.-R.
Sun, JHEP 0905, 061 (2009) [arXiv:0810.0967].

\bibitem{Fursaev} D.V. Fursaev, JHEP 0609, 018 (2006) [hep-th/0606184]; D.V. Fursaev, Phys. Rev. D 77, 124002 (2008) [arXiv:0711.1221].

{\bibitem{Nelson} W. Nelson, Phys. Rev. D 50, 7400-7402 (1994)
[hep-th/9406011].}

{\bibitem{Fursaev2} D.V. Fursaev, Phys. Rev. D 82, 064013 (2010)
[arXiv:1006.2623].}

\bibitem{Bekenstein} J.D. Bekenstein, Phys. Rev. D 23 (1981) 287.

\end{thebibliography}
\end{document}